\documentclass[12pt]{article}
\usepackage{amsmath,amssymb,amsfonts}
\usepackage[paper=letterpaper,margin=1.0in]{geometry}
\usepackage{graphicx}
\usepackage{bm}
\usepackage{color}
\usepackage{setspace}

\begin{document}

%\rightline{\footnotesize Preprint Number}

\begin{center}
\renewcommand{\thefootnote}{\fnsymbol{footnote}}
\vskip 1.0cm
\centerline{\Large\bf 
Echoes of Love Beyond the Horizon: 
}
\centerline{\Large\bf A Bridge to Recovering Information from Black Holes${}\footnote{This essay received an Honorable Mention in the 2025 Essay Competition of the Gravity Research Foundation.}$}
\vskip 1.0cm

\renewcommand{\thefootnote}{\fnsymbol{footnote}}
\centerline{\bf
Meysam Motaharfar${}$\footnote{Corresponding author}${}$\footnote{mmotah4@lsu.edu} and
Parampreet Singh ${}$\footnote{psingh@lsu.edu}
}
\renewcommand{\thefootnote}{\arabic{footnote}}

{\it
${}^1$Department of Physics and Astronomy, Baton Rouge, LA 70803, USA
}

\vskip 1cm

\begin{abstract} 

We provide further evidence that information is preserved during black hole evaporation and may be recoverable, provided quantum gravitational effects resolve the singularity. We demonstrate that due to quantum gravity effects, black holes acquire quantum hair, manifested by non-zero tidal Love numbers, revealing a distinct internal structure similar to neutron stars. Interestingly, the magnitude of these Love numbers is Planck-scale suppressed, implying significant tidal deformation in the late stage of evaporation. Depending on the final state of the black hole, information may be retrieved through correlations in Hawking radiation, baby universes, or via remnants.
    
\end{abstract}

\end{center}

%%%%%%%%%%%%%%%%%%%%%%%%%%%%%%%%%%%%%%%%%%%%%%%%%%%%%%%%%%%%%%%%%%%%%
\newpage

\textbf{Information is conserved.} In a universe governed by deterministic and reversible physical laws, every state of a physical system uniquely maps to its past and future, ensuring the conservation of information. Yet, the black hole evaporation, a seemingly irreversible process, raises a fundamental question: do black holes really destroy information? Classical theory fails to provide a complete description of black holes, primarily because of the presence of singularities. Thus, any answers based on Einstein's theory of General Relativity, even with quantum matter included but without quantizing gravity, would only provide an incomplete description. In this essay, we argue that while classical black holes are featureless, they acquire quantum hair, characterized by non-zero tidal Love numbers, when quantum gravitational effects resolve the singularity. This indicates that quantum black holes possess an internal structure\footnote{By ``internal structure," we refer to the equation of state of matter leading to the black hole.} similar to other astrophysical objects, such as neutron stars, suggesting that the information is indeed preserved and potentially recoverable. Our findings strengthen the idea that the resolution of the information loss paradox is intrinsically linked to the quantum gravitational resolution of singularities. To explore this further, we first refine the formulation of the paradox to uncover a path toward its potential resolution.

\textbf{A classical black hole has no hair.} Consider a mass distribution with total mass $M$ (hereafter referred to as matter $A$), collapsing under its own gravitational field. As the collapse progresses, the entire mass distribution disappears behind the event horizon, forming a black hole. According to the no-hair theorem (hair being a metaphor here for all complicated details) \cite{no-hair}, isolated black holes are remarkably simple, uniquely defined by only a few external macroscopic parameters: mass, angular momentum, and electric charge. All other information about the matter that formed the black hole or any additional matter falling into it is lost behind the horizon and becomes permanently inaccessible to external observers. At this point, some argue that information is truly lost, while others assert that the information is preserved but hidden behind the horizon. The true paradox, however, arises when one considers another remarkable property of black holes.

\textbf{Black holes are not black.} Hawking’s groundbreaking discovery \cite{Hawking:1974rv} surprisingly reveals that black holes are not black in reality and they radiate energy and eventually evaporate. In simple words, the gravitational field near the horizon destabilizes the vacuum, leading to a pair particle production in a process similar to the Schwinger effect. Due to momentum conservation, a particle\footnote{We clarify that by ``particle," we refer to the quantum field modes from a field-theoretic view.} carrying positive energy (hereafter referred to as particle $B$) escapes to infinity, forming Hawking radiation, while its partner (hereafter referred to as particle C), carrying negative energy, falls into the black hole \cite{Hawking:1975vcx}. Then, the black hole gradually loses its mass and energy and starts evaporating. 

\textbf{Who knows what?} To understand the paradox, let us examine the relationship between matter $A$ and particles $B$ and $C$. Since particles $B$ and $C$ are produced from a vacuum, they are maximally entangled and contain correlated information about each other. For instance, if particle $B$ is an electron with spin up, particle $C$ is a positron with spin down. On the other hand, these particles do not encode any information about the initial mass distribution of the collapsing matter $A$ that formed the black hole \cite{Mathur:2009hf}. This arises because the vacuum state near the black hole horizon is uniquely determined by the so-called Unruh vacuum, a state that appears thermal to an accelerating observer, independent of the initial configuration of matter $A$ \cite{Lambert:2013uaa, Hollands:2014eia}. This ties to how the vacuum is defined in curved spacetime, where there is no globally well-defined inertial frame and energy conservation. In fact, a vacuum can uniquely be defined in curved spacetime by requiring the presence of a time-like Killing vector, which ensures that the spacetime remains stationary due to time translation symmetry. However, according to the no-hair theorem, the geometry is uniquely determined by only three macroscopic parameters. Therefore, the associated time-like Killing vector is identical for all mass distributions with the same total mass, charge, and angular momentum, and the vacuum is uniquely determined by the Unruh vacuum, leading to the identical Hawking radiation. 

\textbf{Why a paradox?} As the black hole completely evaporates, only particles $B$ remain, while we started with matter $A$ before black hole formation. This is the so-called black hole information loss paradox \cite{Unruh:2017uaw}, which can be framed in two distinct ways:

\textit{Paradox 1.} Different gravitational collapse scenarios lead to a unique (Unruh) vacuum at the horizon.

\textit{Paradox 2.} A pure initial state evolves to a mixed state.

The first paradox, which clearly highlights the loss of information about the initial mass distribution, has a partly classical origin due to the no-hair theorem. However, it becomes a true paradox when quantum effects are considered, as the Hawking radiation turns out to be independent of the initial mass distribution. The second paradox, which reveals the loss of information through a non-unitary evolution forbidden by quantum mechanics, is entirely a quantum phenomenon. At this stage, we want to emphasize that purity of the final state is necessary but not sufficient for complete resolution of the paradox. To fully recover information, particles $B$ and $C$ should also be able to encode information about the initial mass distribution of matter $A$, requiring violation of the no-hair theorem and necessitating a deviation from the Unruh vacuum at the horizon.

\textbf{Love Numbers.} The original no-hair theorem strictly applies to isolated black holes, a condition that is rarely met in realistic astrophysical scenarios, such as when a black hole is part of a binary system or when it is surrounded by an accretion disk. Yet, intriguingly, black hole simplicity reemerges in a different form. To understand this behavior, it is relevant to examine how a self-gravitating object is deformed in response to the gravitational field of its companion, analogous to material polarizability in electrodynamics. The deformation of black holes and neutron stars is characterized by the so-called tidal Love numbers \cite{10.1093/mnras/69.6.476} of the first and second kind, namely $h_{r}$ and $k_{r}$. Roughly speaking, while $h_{r}$ measures the changes in the shape of the objects, $k_{r}$ quantifies the modification of the asymptotic multipole moments in the presence of an external tidal field.

\textbf{Classical black holes fall in Love.} For a static black hole in the astrophysical environment, it turns out that $h_{r}$ is non-zero due to a change in the geometry of the horizon, while $k_{r}$ exactly vanishes\footnote{Since the astrophysical version of the no-hair theorem has been proven only for Schwarzschild black holes in full General Relativity, we hereafter use the term ``black hole" to specifically refer to the Schwarzschild black holes. However, it is important to note that several families of asymptotically flat black holes in four-dimensional spacetime have exactly vanishing tidal Love numbers within the framework of linear perturbation theory \cite{Love numbers}.} \cite{Gurlebeck:2015xpa}. At first glance, it might seem counterintuitive that the horizon geometry changes without inducing asymptotic multipole moments. This is analogous to Newtonian gravity, where a point-like mass, despite distortions in its equipotential surfaces due to the gravitational field of its companion, remains characterized solely by its monopole moment, highlighting that it lacks any internal structure. Put differently, an object experiences tidal forces from the gravitational field of its companion, with the strength of the force scaling with the size of the object. Similarly, black holes exhibit no tidal response and remain characterized solely by their mass monopole at large distances, retaining their simplicity even when they interact gravitationally. Hence, a classical black hole ``fall in love" with its companion; it remains rigid, revealing nothing about its interior. 

\textbf{The paradox of Love.} From the perspective of General Relativity, two compact objects with the same total mass but different equations of state exhibit distinct tidal responses, allowing an external observer to distinguish their internal structures. However, once these objects collapse and form a black hole, their tidal Love numbers vanish, meaning they no longer exhibit any tidal response. Consequently, they are both solely described by a monopole mass, rendering them structureless and indistinguishable. This suggests a new formulation of the information loss paradox in realistic astrophysical environments:

\textit{Paradox 3.} The Love numbers of various mass distributions undergoing gravitational collapse vanish once they form a black hole.

This new formulation of the paradox is weaker than paradox 1 in the sense that the Hawking radiation can be different when black holes, with identical external macroscopic parameters, are in distinct environments. However, distant observers can measure the tidal Love numbers and confirm that any variation in the Hawking radiation arises solely from changes in the horizon geometry ($h_{r} \neq 0$) due to the surrounding environment, rather than from the initial mass distribution that formed the black holes. In other words, although the Hawking radiation is different, it does not carry any information about the internal structures of black holes due to vanishing tidal Love numbers ($k_{r}=0$). Since we are interested in correlations in the Hawking radiation that encode information about the initial mass distribution, we, hereafter, neglect the effects of the environment on the Hawking radiation for simplicity and assume identical environments.

\textbf{Love is tuned.} From the perspective of effective quantum field theory, the vanishing of all Love numbers for black holes manifests itself as a fine-tuned condition. Therefore, unless a fundamental symmetry protects black hole tidal Love numbers from quantum corrections, their exact vanishing values pose a theoretical puzzle. This suggests that the gravitational potential should receive quantum gravitational corrections, deviating from the classical $1/r$ behavior, similar to how the electrostatic potential receives radiative corrections \cite{Porto:2016zng, Calmet:2021stu}. These corrections lead to non-vanishing Love numbers, which reveal information about the initial mass distribution.

\textbf{Quantum hair.} It is expected that the gravitational field is quantized in the theory of quantum gravity. Specifically, as the spacetime curvature reaches the Planckian scale, the spacetime becomes discrete, and the continuous classical differential geometry is replaced by a quantum discrete geometry. Several studies on black holes in theories of quantum gravity reveal that as the spacetime curvature reaches the Planck scale, an effective repulsive force emerges due to spacetime discreteness. This force, analogous to electron and neutron degeneracy pressure in astrophysical objects, counterbalances gravitational collapse and leads to singularity resolution. For instance, many loop quantum black holes predict that interior singularity is replaced with a finite transition surface that connects the past trapped surface (black hole) to the future anti-trapped surface (white hole) \cite{Ashtekar:2018lag}. Since the matter inside the black hole does not collapse to a singularity but rather reaches a finite size, we anticipate that quantized black holes experience tidal forces and deform when quantum gravitational effects resolve the singularity. This indicates the presence of a quantum hair, characterized by non-zero tidal Love numbers, through which black holes reveal their distinct internal structure similar to neutron stars. 

For a matter collapsing under its own gravitational field, quantum gravity becomes important when spacetime curvature becomes Planckian, i.e., $\rho \sim \rho_{\textrm{Pl}}$, or when the radius is $r_{\textrm{QG}} \sim \left({M}/{M_{\textrm{Pl}}}\right)^{1/3} l_{\textrm{Pl}}$, with $l_{\textrm{Pl}}  = \sqrt{\hbar G/c^3}$ being the Planck length. However, such quantum gravitational effects should be suppressed at the horizon, implying that quantum gravitational correction to tidal Love numbers is Planck-scale suppressed, i.e., $k_{r} \propto r_{\textrm{QG}}/r_{\textrm{S}} \sim (M_{\textrm{Pl}}/M)^{2/3}$, with $r_{S} = 2GM/c^2$ and $M_{\textrm{Pl}} = \sqrt{\hbar c/G}$ being the Schwarzschild radius and Planck mass, respectively \cite{Motaharfar:2025typ, second draft, Kim:2020dif}. Interestingly, the tidal deformation becomes stronger as the black hole mass reaches Planck mass, where Hawking radiation also becomes significant. 

\textbf{Unitary evolution is not sufficient.} There are two well-studied possible scenarios for the final state of black holes in theories of quantum gravity that predict a discrete spacetime at the Planck scale. Black holes either transit to a white hole (baby universe) or leave a remnant as they evaporate. In both cases, it is claimed that the information loss paradox is resolved \cite{Ashtekar:2020ifw, Ashtekar:2025ptw, Ashtekar:2008jd, Rovelli:2024sjl}. Let us examine the first scenario where the black holes are isolated and the interior singularity of the black hole is regularized due to quantum gravity. Consider again the mass distribution $A$ collapsing under its own gravitational field. Once it becomes a black hole, it is only defined by the total mass $M$ from a distant observer's perspective. However, in this case, as the energy density of matter $A$ inside the horizon reaches Planck density, effective repulsive forces resolve singularity and allow for a black hole to white hole transition. Consequently, one expects that particles $C$, which fall into the black hole, will later emerge from the white hole, thereby purifying the Hawking radiation. \cite{Ashtekar:2020ifw, Ashtekar:2025ptw}. Although a complete underlying mechanism is still unknown due to the absence of a full theory of quantum gravity, the purification of the Hawking radiation only guarantees the unitary evolution but not a complete information recovery, as paradox 1 persists as long as the no-hair theorem holds.

\textbf{Love recovers information.} Let us consider quantum gravitational black holes in astrophysical scenarios. In this case, the black hole formed from different mass distributions is deformed in a distinct way. The geometry of the black hole is different at the horizon, leading to different vacuums for different mass distributions (remember that we are assuming an identical environment so the change of geometry merely comes from the difference in mass distribution). Therefore, when particles $C$ come out of the white hole and purify Hawking radiation, they have information about the initial mass distribution. Even if the correlation is extremely small, of the order of $e^{-S_{\textrm{BH}}}$, with $S_{\textrm{BH}}$ being the entropy of the black hole, it is sufficient to purify the mixed state, ensuring the unitary evolution and information recovery. This follows from the fact that, independent of any specific dynamical assumptions, only a purely kinematic analysis reveals that random pure states in a system with a large number of degrees of freedom, such as $e^{S_{\textrm{BH}}}$ in the case of black holes, are exponentially close to mixed states \cite{Raju:2020smc}. In our case, the leading order corrections to Love numbers are of order $S^{-1/3}_{\textrm{BH}}$. Therefore, we expect a correlation of the same order, which is even larger than $e^{-S_{\textrm{BH}}}$, ensuring the information recovery. Thus, singularity resolution not only restores unitarity evolution but also provides a concrete mechanism for information retrieval.

\textbf{Remnants are distinguishable.} Finally, consider the scenario in which a black hole evaporates to the point where its energy density approaches the Planck scale. At this stage, quantum gravitational effects become dominant, potentially halting further radiation and leaving behind a stable remnant \cite{Rovelli:2024sjl}. Such a remnant could serve as a natural resolution to the information loss paradox, preserving the information that would otherwise be lost in the classical picture of black hole evaporation. However, it is argued that since potentially there are large numbers of remnants, due to various collapse scenarios, of order $e^{S_{\textrm{BH}}}$, remnants should be overproduced even though the production rate of any specific remnant from the quantum vacuum is exponentially suppressed, i.e., $e^{-S_{\textrm{BH}}}$ \cite{Chen:2014jwq}. This is true for the case that one does not take into account the non-vanishing tidal Love numbers of quantum black holes. In that case, not only the information leak outside the black hole due to correlation in the Hawking radiation but also the remnant is gravitationally interacting, revealing its distinct internal structure through tidal response and breaking its degeneracy. Furthermore, remnants are no longer indistinguishable point-like particles since they have Planckian mass and are significantly deformed by tidal forces. Thus, the overproduction argument based on point-like effective field theory does not apply indeed. This suggests that the non-vanishing tidal Love numbers provide further support for remnants as a possible resolution of the information loss paradox in theories of quantum gravity. 

\textbf{The challenge is to listen.} Our findings reinforce the idea that there is no information loss paradox if black hole singularities are resolved by quantum gravitational effects. Specifically, quantum gravitational corrections to Love numbers, even if small, suggest that black holes may whisper their deepest secrets in the ripples of spacetime. The task is to tune in and decipher their message.

\noindent
\textbf{Acknowledgments:
} 
We thank Maxwell R. Siebersma for discussions. This work is supported by NSF grants PHY-2110207
and PHY-2409543.


\begin{thebibliography}{9}

\bibitem{no-hair}
W.~Israel,
``Event horizons in static vacuum space-times,''
Phys. Rev. \textbf{164}, 1776-1779 (1967);
B.~Carter,
``Axisymmetric Black Hole Has Only Two Degrees of Freedom,''
Phys. Rev. Lett. \textbf{26}, 331-333 (1971);
S.~W.~Hawking,
``Black holes in general relativity,''
Commun. Math. Phys. \textbf{25}, 152-166 (1972).

%\cite{Hawking:1974rv}
\bibitem{Hawking:1974rv}
S.~W.~Hawking,
``Black hole explosions,''
Nature \textbf{248}, 30-31 (1974).
%doi:10.1038/248030a0
%4923 citations counted in INSPIRE as of 19 Mar 2025

%\cite{Hawking:1975vcx}
\bibitem{Hawking:1975vcx}
S.~W.~Hawking,
``Particle Creation by Black Holes,''
Commun. Math. Phys. \textbf{43}, 199-220 (1975).
%[erratum: Commun. Math. Phys. \textbf{46}, 206 (1976)]
%doi:10.1007/BF02345020
%12065 citations counted in INSPIRE as of 27 Mar 2025

%\cite{Mathur:2009hf}
\bibitem{Mathur:2009hf}
S.~D.~Mathur,
``The Information paradox: A Pedagogical introduction,''
Class. Quant. Grav. \textbf{26}, 224001 (2009).
%doi:10.1088/0264-9381/26/22/224001
%[arXiv:0909.1038 [hep-th]].
%916 citations counted in INSPIRE as of 27 Mar 2025

%\cite{Lambert:2013uaa}
\bibitem{Lambert:2013uaa}
P.~H.~Lambert,
``Introduction to Black Hole Evaporation,''
PoS \textbf{Modave2013}, 001 (2013).
%doi:10.22323/1.201.0001
%[arXiv:1310.8312 [gr-qc]].
%13 citations counted in INSPIRE as of 27 Mar 2025
%\cite{Hollands:2014eia}

\bibitem{Hollands:2014eia}
S.~Hollands and R.~M.~Wald,
``Quantum fields in curved spacetime,''
Phys. Rept. \textbf{574}, 1-35 (2015).
%doi:10.1016/j.physrep.2015.02.001
%[arXiv:1401.2026 [gr-qc]].
%238 citations counted in INSPIRE as of 27 Mar 2025

%\cite{Unruh:2017uaw}
\bibitem{Unruh:2017uaw}
W.~G.~Unruh and R.~M.~Wald,
``Information Loss,''
Rept. Prog. Phys. \textbf{80}, no.9, 092002 (2017).
%doi:10.1088/1361-6633/aa778e
%[arXiv:1703.02140 [hep-th]].
%244 citations counted in INSPIRE as of 27 Mar 2025

\bibitem{10.1093/mnras/69.6.476}
A. E. H. Love,
``The yielding of the Earth to disturbing forces,''
Monthly Notices of the Royal Astronomical Society \textbf{69}, 476 (1909).

%\cite{Gurlebeck:2015xpa}
\bibitem{Gurlebeck:2015xpa}
N.~G\"urlebeck,
``No-hair theorem for Black Holes in Astrophysical Environments,''
Phys. Rev. Lett. \textbf{114}, no.15, 151102 (2015).

\bibitem{Love numbers}
A.~Le Tiec and M.~Casals,
``Spinning Black Holes Fall in Love,''
Phys. Rev. Lett. \textbf{126}, no.13, 131102 (2021);
A.~Le Tiec, M.~Casals and E.~Franzin,
``Tidal Love Numbers of Kerr Black Holes,''
Phys. Rev. D \textbf{103}, no.8, 084021 (2021);
H.~S.~Chia,``Tidal deformation and dissipation of rotating black holes,''
Phys. Rev. D \textbf{104}, no.2, 024013 (2021);
P.~Charalambous, S.~Dubovsky and M.~M.~Ivanov,
``On the Vanishing of Love Numbers for Kerr Black Holes,''
JHEP \textbf{05}, 038 (2021); R.~P.~Bhatt, S.~Chakraborty and S.~Bose, ``Response of a Kerr black hole to a generic tidal perturbation,''
[arXiv:2412.15117 [gr-qc]]; R.~P.~Bhatt, S.~Chakraborty and S.~Bose,
``Rotating black holes experience dynamical tides,''
[arXiv:2406.09543 [gr-qc]]; D.~Pere\~niguez and V.~Cardoso,
``Love numbers and magnetic susceptibility of charged black holes,''
Phys. Rev. D \textbf{105}, no.4, 044026 (2022); M.~Rai and L.~Santoni,
``Ladder symmetries and Love numbers of Reissner-Nordstr\"om black holes,'' JHEP \textbf{07}, 098 (2024); L.~Ma, Z.~H.~Wu, Y.~Pang and H.~Lu,
``Charging the Love numbers: Charged scalar response coefficients of Kerr-Newman black holes,''
[arXiv:2408.10352 [gr-qc]].

%\cite{Porto:2016zng}
\bibitem{Porto:2016zng}
R.~A.~Porto,
``The Tune of Love and the Nature(ness) of Spacetime,''
Fortsch. Phys. \textbf{64}, no.10, 723-729 (2016).
%doi:10.1002/prop.201600064
%[arXiv:1606.08895 [gr-qc]].
%129 citations counted in INSPIRE as of 18 Mar 2025

%\cite{Calmet:2021stu}
\bibitem{Calmet:2021stu}
X.~Calmet, R.~Casadio, S.~D.~H.~Hsu and F.~Kuipers,
``Quantum Hair from Gravity,''
Phys. Rev. Lett. \textbf{128}, no.11, 111301 (2022).
%doi:10.1103/PhysRevLett.128.111301
%[arXiv:2110.09386 [hep-th]].
%37 citations counted in INSPIRE as of 18 Mar 2025

%\cite{Ashtekar:2018lag}
\bibitem{Ashtekar:2018lag}
A.~Ashtekar, J.~Olmedo and P.~Singh,
``Quantum Transfiguration of Kruskal Black Holes,''
Phys. Rev. Lett. \textbf{121}, no.24, 241301 (2018).
%doi:10.1103/PhysRevLett.121.241301
%[arXiv:1806.00648 [gr-qc]].
%239 citations counted in INSPIRE as of 18 Mar 2025

%\cite{Motaharfar:2025typ}
\bibitem{Motaharfar:2025typ}
M.~Motaharfar and P.~Singh,
``Loop Quantum Gravitational Signatures via Love Numbers,''
[arXiv:2501.09151 [gr-qc]].
%1 citations counted in INSPIRE as of 19 Mar 2025

\bibitem{second draft}
M. Motaharfar and P. Singh, ``Love Numbers of Covariant Loop Quantum Black Holes," [arXiv:2505.14784 [gr-qc]].

%\cite{Kim:2020dif}
\bibitem{Kim:2020dif}
J.~W.~Kim and M.~Shim,
``Quantum corrections to tidal Love number for Schwarzschild black holes,''
Phys. Rev. D \textbf{104}, no.4, 046022 (2021)
%doi:10.1103/PhysRevD.104.046022
%[arXiv:2011.03337 [hep-th]].
%18 citations counted in INSPIRE as of 28 Mar 2025

%\cite{Ashtekar:2020ifw}
\bibitem{Ashtekar:2020ifw}
A.~Ashtekar,
``Black Hole evaporation: A Perspective from Loop Quantum Gravity,''
Universe \textbf{6}, no.2, 21 (2020).
%doi:10.3390/universe6020021
%[arXiv:2001.08833 [gr-qc]].
%71 citations counted in INSPIRE as of 28 Mar 2025

%\cite{Ashtekar:2025ptw}
\bibitem{Ashtekar:2025ptw}
A.~Ashtekar,
``Black hole evaporation in loop quantum gravity,''
Gen. Rel. Grav. \textbf{57}, no.2, 48 (2025).
%doi:10.1007/s10714-025-03380-7
%[arXiv:2502.04252 [gr-qc]].
%4 citations counted in INSPIRE as of 28 Mar 2025

%\cite{Ashtekar:2008jd}
\bibitem{Ashtekar:2008jd}
A.~Ashtekar, V.~Taveras and M.~Varadarajan,
``Information is Not Lost in the Evaporation of 2-dimensional Black Holes,''
Phys. Rev. Lett. \textbf{100}, 211302 (2008).
%doi:10.1103/PhysRevLett.100.211302
%[arXiv:0801.1811 [gr-qc]].
%105 citations counted in INSPIRE as of 28 Mar 2025

%\cite{Rovelli:2024sjl}
\bibitem{Rovelli:2024sjl}
C.~Rovelli and F.~Vidotto,
``Planck stars, White Holes, Remnants and Planck-mass quasi-particles. The quantum gravity phase in black holes' evolution and its manifestations,''
[arXiv:2407.09584 [gr-qc]].
%13 citations counted in INSPIRE as of 19 Mar 2025

%\cite{Raju:2020smc}
\bibitem{Raju:2020smc}
S.~Raju,
``Lessons from the information paradox,''
Phys. Rept. \textbf{943}, 1-80 (2022).
%doi:10.1016/j.physrep.2021.10.001
%[arXiv:2012.05770 [hep-th]].
%212 citations counted in INSPIRE as of 18 Mar 2025


%\cite{Chen:2014jwq}
\bibitem{Chen:2014jwq}
P.~Chen, Y.~C.~Ong and D.~h.~Yeom,
``Black Hole Remnants and the Information Loss Paradox,''
Phys. Rept. \textbf{603}, 1-45 (2015).
%doi:10.1016/j.physrep.2015.10.007
%[arXiv:1412.8366 [gr-qc]].
%307 citations counted in INSPIRE as of 19 Mar 2025


\end{thebibliography}
\end{document}